\documentclass[aps,twocolumn,floatfix]{revtex4}
\usepackage{graphicx}
\usepackage{epsfig}
\def \CRR {{\rm Cr{(Ru}_2)}_3}

\def \TC {T_{\rm c}}
\def \vH {{\bf H}}

\def \Jc {J_c}
\def \Kc {K_c}

\def \vm {{\bf m}}

\def \vmag {{\bf M}}
\def \mavg {M_{av}}
\def \ncr {{N_{\rm Cr}}}

\def \msat {M_{sat}}
\def \hsf {H_{sf}}
\def \msl {M_{sl}}
\def \vno {{\bf n}_1}
\def \vnt {{\bf n}_2}

\def \vnoi {{\bf n}_{1i}}
\def \vnti {{\bf n}_{2i}}
\def \vnj {{\bf n}_j}
\def \chsl {\chi_{sl}} 
\def \chn {\chi_{nint}}

\begin{document}

\date{\today}
\title{Giant Antiferromagnetically Coupled Moments in a Molecule-Based Magnet with Interpenetrating Lattices}
\author{Randy S. Fishman$^1$, Satoshi Okamoto$^1$, William W. Shum$^2$, and Joel S. Miller$^2$}
\affiliation{$^1$Materials Science and Technology Division, Oak Ridge National Laboratory, 
Oak Ridge, Tennessee 37831-6071}
\affiliation{$^2$Department of Chemistry, University of Utah, Salt Lake City, Utah 84112-0850}

\begin{abstract}

The molecule-based magnet [Ru$_2$(O$_2$CMe)$_4$]$_3$[Cr(CN)$_6$] contains two
weakly-coupled, interpenetrating sublattices in a body-centered cubic structure.  Although the 
field-dependent magnetization indicates a metamagnetic transition from an antiferromagnet to a paramagnet, 
the hysteresis loop also exhibits a substantial magnetic remanance and coercive field uncharacteristic of a 
typical metamagnet.   We demonstrate that this material behaves like two giant moments with a weak antiferromagnetic 
coupling and a large energy barrier between the orientations of each moment.   Because the sublattice moments
only weakly depend on field in the transition region, the magnetic correlation length can be directly estimated
from the magnetization.

\end{abstract}
\maketitle

\newpage

Due to the importance of weak interactions that are commonly neglected in solid-state materials,
molecule-based magnets exhibit a variety of novel behavior \cite{Miller94, Blundell04}.
One of the most fascinating and puzzling molecule-based magnets containing two
interpenetrating lattices of [Ru$_2$(O$_2$CMe)$_4$]$_3$[Cr(CN)$_6$] (Me = methyl, CH$_3$) \cite{Liao02, Vos04}
($\CRR $) exhibits an unusual ``wasp-waisted" hysteresis loop that is attributed
to the weak antiferromagnetic (AF) coupling between its two sublattices \cite{Vos05, Miller05}.
The shape of the initial $M(H)$ data plotted in Fig.1 \cite{Shum08}, with a shallow slope followed by a rapid rise, 
is commonly associated with a metamagnetic transition between AF and paramagnetic (PM) states \cite{Str77}, 
which has been observed both in solid-state systems like FeCl$_2$ \cite{FeCl} and DyAl garnet \cite{DyAl}
as well as in several other molecule-based magnets \cite{Molb}.  However, typical 
metamagnets do not exhibit the nonzero remanent magnetization and coercive field 
found in the hysteresis loop of $\CRR $ compounds with interpenetrating lattices.  

We argue that a $\CRR $ compound with interpenetrating lattices behaves like two 
macroscopic AF coupled moments with a large energy barrier between the orientations 
of each moment.  Quasi-one- or two-dimensional materials that magnetically order in three dimensions 
due to a weak coupling between sublattices are fairly common, even among molecule-based magnets 
\cite{Coronado05}.  Much rarer are materials where each of the weakly-interacting sublattices 
is fully ordered in three dimensions.

\begin{figure}
\includegraphics *[scale=0.5]{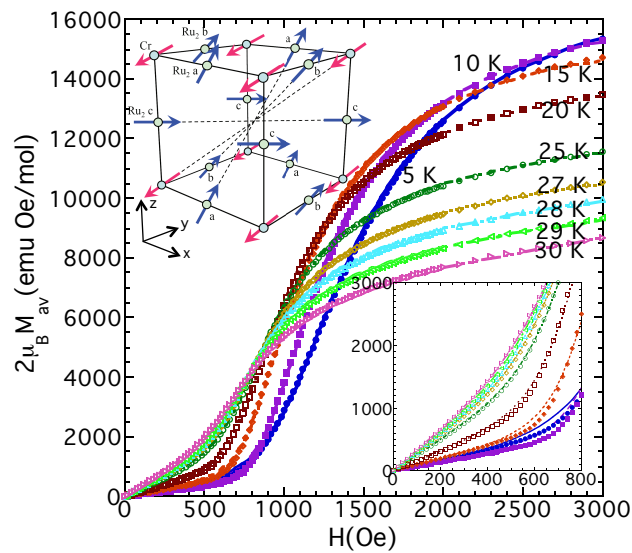}
\caption{
The initial magnetization (virgin curve) of $\CRR $ with interpenetrating sublattices for temperatures up to 30 K along with the 
predicted field dependence.  The upper left inset shows the predicted ground state of a single sublattice for classical spins
with large anisotropy.}
\end{figure}

A single lattice of $\CRR $ is sketched in the inset to Fig.1, where every pair of Ru ions bridges two [Cr(CN)$_6$]$^{3-}$ 
(Cr) ions located at the corners of the cubic unit cell and separated by $a_l= 13.4$ \AA .  
While each Cr(III) ion has a  spin $S=3/2$, each Ru$_2$ dimer is in a II/III mixed-valence state 
with a total spin $S=3/2$ \cite{Liao02}.  Due to the ``paddle-wheel" molecular environment produced by the 
surrounding four Me groups, each Ru$_2$ moment experiences a strong easy-plane anisotropy with 
$D\approx 100$ K or 8.6 meV \cite{Misk99, Shum04}.  The easy plane lies perpendicular to the 
axis joining the Ru$_2$ sites with the neighboring Cr ions.  Although a single-lattice 
$\CRR $ compound has been synthesized \cite{Vos05}, the resulting samples are amorphous.  
A $\CRR $ compound with interpenetrating lattices contains a second identical lattice inserted through 
the open space of the first lattice resulting in a body-centered cubic structure.

Most of the properties of the interpenetrating-lattice compound can be explained by 
a simple model with strong easy-plane anisotropy $D$ on the Ru$_2$ sites, AF
intra-sublattice exchange $\Jc $ between neighboring Cr and Ru$_2$ sites on each sublattice, and a 
weak AF inter-sublattice exchange $\Kc$ between moments on the two sublattices.  
The coupling $\Kc $ is the sum of the dipolar energy and the superexchange interaction through 
weakly-overlapping molecular orbitals on the two sublattices \cite{dip}.

The key to understanding $\CRR $ with interpenetrating lattices is to construct the correct 
ground state for the single-lattice compound.  Each Ru$_2$ spin pair in Fig.1 is labeled
as $a$ (along the $x$ axis), $b$ (along $y$), or $c$ (along $z$).  In the classical limit with 
infinite anisotropy, the $a$, $b$, or $c$ spins must lie the $yz$, $xz$, or $xy$ planes, respectively. 
In the quantum case, the Ru$_2$ spins will have small, but 
nonzero components in the classically-forbidden directions.  
For both classical and quantum spins, AF order is frustrated by the 
easy-plane anisotropy.  A similar situation arises in cubic pyrochlores like Ho$_2$Ti$_2$O$_7$, 
where ferromagnetic order is frustrated by local [111] anisotropy \cite{Harris97}.

Both classical and quantum calculations provide the same magnetic ground state.  The sum of the 
Ru$_2$ $a$, $b$, and $c$ spins points opposite to the Cr spin along one of the four diagonals of the cube.  
Accounting for the two orientations of the moment along each diagonal,
there are eight domains in zero field.   For classical spins with infinite anisotropy, the net moment
along one of the diagonals is $\msl  = (\sqrt{6}-1)S \approx 1.45 S$ per Cr ion.    
In both classical and quantum calculations, the Ru$_2$ spins approach the cubic diagonal 
as $D/\Jc $ is reduced.  The quantum result $\msl \approx 1.21 S$ for the single-lattice moment 
with $D/\Jc =5$ is lower than the classical result because the expectation values 
of the Ru$_2$ spins are reduced in magnitude from 1.5 to about 1.23. 
 
Based on quantum mean-field calculations with $D/\Jc =5$ and $\Jc =1.72$ meV (see below),
the single-lattice compound experiences a spin-flop transition at 
$\hsf \approx 2.78 \Jc \mu_B \approx 80$ T with
the field along one of the cubic diagonals.  Above $\hsf $, the 
Cr and total Ru$_2$ moments cant away from the cubic diagonal and, until all the moments
become aligned ferromagnetically (FM) at an even higher field, the magnetization $2\mu_B \vmag $ is 
no longer parallel to $\vH $.  

Now consider $\CRR $ with interpenetrating lattices.  Assuming that
$\Kc \ll \Jc$ and $\mu_B H\ll \Jc S$, the sublattice moments $\msl {\bf n}_j$ ($j$ = 1 or 2)
are rigid with the same spin configuration as in the inset to Fig.1 but pointing along arbitrary 
cubic diagonals.  Thermal equilibrium between the 64 possible configurations $\{ \vno ,\vnt \}$
is established within correlated clusters of size $\xi $ containing $\ncr \sim 2(\xi /a_l)^3$ Cr
spins per cluster.   Although $\xi $ depends on field and may peak in the transition region, the polycrystalline 
nature of the $\CRR $ samples suggests that the average value of $\xi $ is most important.  
With $\ncr $ assumed independent of field, the energy is given by
\begin{eqnarray}
\label{Enc}
E&= &\ncr \sum_i \Bigl\{ -\mu_B \msl (\vnoi +\vnti )\cdot \vH \nonumber \\
&+&3 K_c S^2 \,\vnoi \cdot \vnti \Bigr\} \ ,
\end{eqnarray}
where $\vH =H\vm $ is the magnetic field and we sum over clusters $i$  
The intra-sublattice exchange $\Jc $ only enters $E$ through the 
sublattice moments $\msl (T)$, which vanish above $\TC \propto \Jc S^2$. 

\begin{figure}
\includegraphics *[scale=0.5]{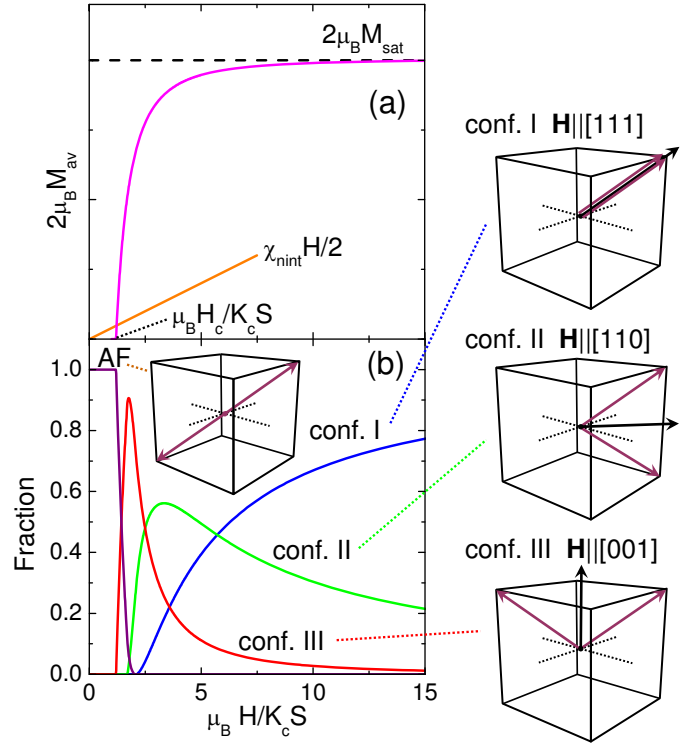}
\caption{
(a) The $T=0$ average moment and non-interacting linear susceptibility $\chn H/2$ 
and (b) the fractions of the AF state and configurations I, II, and III versus 
$\mu_B H/\Kc S$ assuming classical spins and large anisotropy.
}
\end{figure}

The zero-field AF ground state is shown schematically in Fig.2(b) with $\vno =-\vnt $.  
Depending on the field orientation, the AF state becomes unstable to one of three possible spin configurations.  
Configuration III with $\vno \cdot \vnt =-1/3$ appears at the critical field $H_c=\sqrt{3}\Kc S^2/(\mu_B \msl )$
when $\vH $ is parallel to [111].  In configuration II, $\vno\cdot \vnt =1/3$ and in the PM state labeled configuration I, 
$\vno = \vnt $.  The stable phases are shown in 
Fig.2 along with the fraction of configurations I, II, and III when averaged over all field directions $\vm $.
Notice that the AF configuration vanishes only with the appearance of the PM state.  
In zero field, $\mavg $ obeys the simple functional form $\mavg /\msat =1-(H_c/H)^2$ plotted in Fig.2(a).  

We must modify Eq.(\ref{Enc}) to account for the small distortion
of the single-lattice ground state with field.  That distortion is responsible for the small linear susceptibility observed
within the AF state at low temperature and small fields, and for the even smaller differential susceptibility $2\mu_BdM/dH$
observed within the PM state at high fields.  For classical spins and large anisotropy, the magnetic ground state 
with $\vnj =\vm $ cannot be deformed until very large fields, when the Cr spins tilt away from the
$-\vm $ direction.  But for $\vnj =-\vm $, the Ru$_2$ spins can easily cant towards the field direction.  

So the simplest form for the susceptibility of sublattice $j$ is 
$\chi_j = \chsl \sin^2 (\theta_j /2)$ where $\theta_j =\cos^{-1} (\vnj \cdot \vm )$ is the angle 
between the field and moment directions.
Hence,  $\chi_j=0$ when $\theta_j=0$ and $\chi_j = \chsl $ when $\theta_j=\pi $.  
The non-interacting susceptibility of the magnetic configuration $\{ \vno ,\vnt \}$ is then given by
$\chn = \chi_1 +\chi_2$ per pair of Cr atoms and the additional linear term in the magnetization
is $\ncr \chn H/2$.  The extra term $- \ncr \chn H^2/4$ must then be added to the energy $E$ of Eq.(\ref{Enc}) 
for each cluster.

In the AF state with $\vno =-\vnt $, $\theta_2 =\theta_1+\pi $ and $\chn = \chsl $.  In the PM state with
$\theta_1=\theta_2$, $\chn = 2\chsl \sin^2 (\theta_1/2)$ vanishes when $\theta_1=0$
but is nonzero when $\vH $ points away from a cubic diagonal.
Averaged over all field directions, the susceptibility in the PM phase is $\chn \approx 0.134 \chsl $.
So in agreement with experiments \cite{Vos04, Shum08}, the differential susceptibility
at high fields is much smaller than the the linear susceptibility at low fields.  

\begin{figure}
\includegraphics *[scale=0.7]{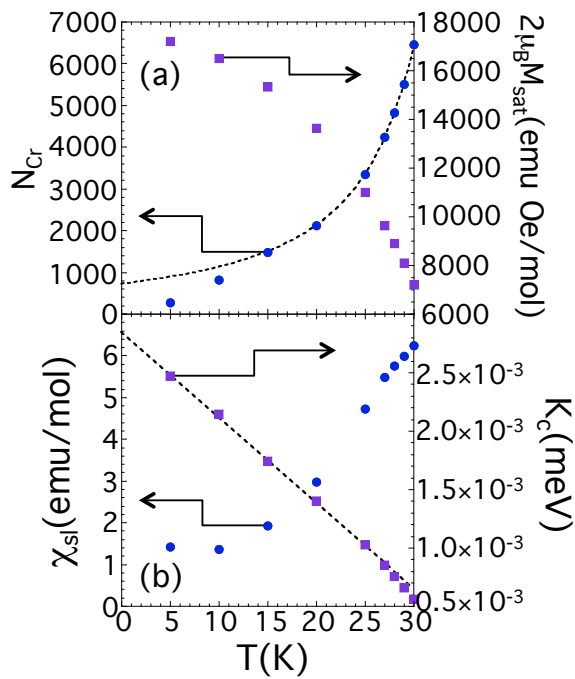}
\caption{
The fitting parameters versus temperature:  (a) the number $\ncr (T)$ of Cr atoms in each 
magnetic cluser (the dashed curve gives the fit $\ncr = N_0[(\TC-T)/\TC]^{-3\nu }$ with $\nu =0.5$) 
and the saturation magnetization $2\mu_B\msat (T)$ and (b) the single-lattice susceptibility $\chsl (T)$ and the 
inter-sublattice interaction $\Kc (T)$.
}
\end{figure}

Using this model, we evaluate the partition function by summing over the
64 configurations of $\{\vno ,\vnt \}$ and the magnetization by averaging over
field directions.  For every temperature, the average magnetization depends on 
four parameters:  the single-lattice susceptibility $\chsl $, the sublattice moment $\msl $, the weak AF interaction
$\Kc $ between sublattices, and the number $\ncr $ of Cr atoms within each cluster, half belonging to 
each sublattice.  The fits are provided in Fig.1 and the resulting parameters are plotted in Fig.3.

In Fig.3(a), $\ncr (T)$ is consistent with the functional form $\ncr (T)= N_0[(\TC-T)/\TC]^{-3\nu }$,
where $N_0 \approx 725 $, $\TC =39$ K,  and $\nu =0.5$, which is the mean-field value for the critical
exponent.  Since the functional form $\xi (T) \propto [(\TC - T)/\TC ]^{-\nu }$ is only expected close to $\TC $, 
it is remarkable that deviations from this form are only apparent at 10 K.  Other evidence for 
the mean-field nature of the phase transition in $\CRR $ comes from the magnetization of the 
single-lattice compound \cite{Vos04}, which although amorphous closely follows a Brillouin function and vanishes 
near $\TC $ like $(\TC -T)^{1/2}$.  

The ``saturation" magnetization
$2\mu_B \msat (T)= \sqrt{3} \mu_B \msl (T)$ per Cr atom plotted in Fig.3(a) gives
the average magnetization when the two nondistorted sublattice moments are aligned.  Nevertheless, 
the magnetization continues to rise with increasing field due to the deformation of the 
ferrimagnetic ground state of each sublattice.   Indeed, Vos {\it et al.} \cite{Vos04} found that the magnetization at 5 K
rises from 16,937 emu Oe/mol at 0.5 T to 20,800 emu Oe/mol at 5 T, far above the ``saturation" value
of 17,286 emu Oe/mol.   In rough agreement with the assumed form for the non-interacting
susceptibility $\chn $, the high-field differential susceptibility of 0.086 emu/mol is about nine times smaller
than the low-field linear susceptibility of 0.81 emu/mol.  Notice that the 30 K clusters with $\ncr = 6464$ have
correlation lengths $\xi \sim 15 a_l=200 \, \AA $ and sublattice magnetizations 
$\mu_B \ncr \msl (T=30\, K)\approx 4500 \mu_B$.

At $T=5$ K, the saturation magnetization is significantly
lower than the classical result with $D=\infty $.  But the single-lattice moment $\msl $ for 
quantum spins decreases rapidly with increasing $D/\Jc $ from its maximal value of $2S$ per
Cr ion when $D=0$.  The fitted saturation value can be used to 
estimate that $D/\Jc \approx 5$ and $\Jc \approx 1.72$ meV.  Another estimate for $\Jc $ comes from
the mean-field result $\TC \approx 4.05 \Jc =33$ K for the transition temperature with $D/\Jc =5$,
yielding the smaller value $\Jc \approx 0.70$ meV.   However, 
mean-field theory typically overestimates $\TC $ and underestimates $\Jc $.   So
at low temperatures, the inter-sublattice coupling $\Kc $ is roughly 600 times smaller than 
the intra-sublattice coupling $\Jc $.

Plotted in Fig.3(b), the sublattice susceptibility $\chsl $
rises sharply with increasing temperature and is roughly proportional to $\ncr $. 
The inter-sublattice coupling $\Kc (T)$ falls off almost linearly with temperature below 25 K.
If $\Kc $ were short ranged and coupled sublattice moments at distinct
points in space, we would expect that $\Kc (T) \propto \msl(T)^2$.  
The temperature dependence of $\Kc (T)/\msl (T)^2$ may be
ascribed to the complex and long-ranged interaction between the two sublattices, each of which contains
two species of magnetic ions with different temperature-dependent average moments.
   
Three of the four fitting parameters in this model, $\msl (T)$, $\ncr (T)$, and $\chsl (T)$,
are properties of the single-lattice compound;  only $\Kc (T)$ reflects the presence of 
two sublattices.  Above about 30 K, the fits break 
down primarily because the linear terms $\chsl (T)\sin^2(\theta /2) H$
in the sublattice magnetization are no longer small compared to the zero-field value of $2\mu_B \msl (T)$.  
Consequently, the inter-sublattice coupling $\Kc $ effectively depends on field as well as on temperature.  

Despite the success of the fits below 30 K, the lower-right inset to Fig.1 suggests one 
limitation of our model.  Although the correlation length $\xi $ must reach a maximum near 
$H_c (T)=\sqrt{3}\Kc(T) S^2/(\mu_B \msl (T))$ as fluctuations soften, our model assumes that
$\xi (T) $ is independent of field.  Consequently, the curves in Fig.1 lie slightly above the data points near $H_c(T)$,
especially at lower temperatures.  A correlation length $\xi (T,H)$ that peaks in the vicinity of $H_c(T)$ would 
produce even better agreement with the experimental data.  
Because the thermal averages are assumed to be independent from one cluster to another, the model magnetization
is a continuous function of $H$ for any nonzero temperature and for any orientation of the magnetic field.
Due to the polycrystalline sample, we cannot say whether the 
experimental magnetization of $\CRR $ experiences a jump near $H_c(T)$ at nonzero temperatures, 
as observed in single crystals of conventional metamagnets \cite{FeCl, DyAl} up to about 85\% of the 
N\'eel temperature.

The interpenetrating-lattice compound $\CRR $ may be the only known material with
two or more weakly-interacting three-dimensional sublattices.  Another molecule-based
magnet that bears some similarities to $\CRR $ is methylamine chrome alums \cite{CAL}, which contains Cr(III)
ions in a face-centered cubic structure.   Although weak AF dipolar interactions couple the four sublattice moments, 
the intra-sublattice exchange is negligible in chrome alums.  So the three-dimensional 
ordering at about 0.02 K is also produced by the dipolar interactions between magnetic sublattices.

Most of the puzzling properties of $\CRR $ with interpenetrating lattices can be explained by two 
macroscopic moments that are weakly AF coupled with a large energy barrier between the 
orientations of each moment.  If the individual moments within each sublattice rotate together, 
then the energy barrier between moments in the [111] and [11${\bar 1}$] directions for classical spins
at $T=0$ is $\Delta(0)=0.07 \Jc S^2\ncr $.  At nonzero temperatures, we expect that 
$\Delta (T)\propto \ncr (T) \msl(T)^2 \Jc \sim N_0 \Jc [(\TC -T)/\TC ]^{-1/2}$.  So $\Delta (T)$ will 
rise very close to $\TC $, which may explain the anomalous AC susceptibility \cite{Liao02} observed 
in these materials.

Our work provides several new predictions for $\CRR $ with interpenetrating lattices.  
In the magnetic ground state, the total moment of each sublattice must lie along one of the four cubic diagonals.
This can be verified by neutron scattering, either on a deuterated sample or with a large enough polycrystal.  
With a single crystal, the transitions between the AF and configurations I, II, or III should be observable
for different orientations of the field.   The predicted correlation length $\xi (T)$ can be verified independently
by fitting the elastic peaks measured with neutron scattering, even on a polycrystalline sample.
We hope that this paper will inspire future work on this fascinating system, including 
measurements on single crystals when they become available.   

{\bf Acknowledgements.}  We would like to acknowledge useful conversations with Prof. Bruce
Gaulin.  This research was sponsored by the Division of Materials Science
and Engineering of the U.S. Department of Energy and by the U.S. National Science Foundation 
(Grant No. 0553573).


\begin{references}
\bibitem{dip} For easy-plane anisotropy,  the Cr-Cr dipolar interactions between sublattices cancel by symmetry while
the Cr-Ru$_2$ and Ru$_2$-Ru$_2$ dipolar interactions produce inter-sublattice interactions that are AF and FM, 
respectively, with a net FM dipolar interaction between
sublattices.  At low temperatures, the observed AF interaction $\Kc $ can be recovered with an additional
inter-sublattice superexchange between neighboring Ru$_2$ pairs separated by $a_l/2$ of
$3.1\times 10^{-3}$ meV, using the sublattice ground state provided by a quantum calculation with
$D/\Jc =5$.
\end{references}
\end{document}